\documentclass[10pt, conference]{IEEEtran}

\usepackage{amsmath}
\usepackage{amsfonts}
\usepackage{amssymb}
\usepackage{amsthm}
\usepackage{tabularx}
\usepackage{mathrsfs}

\usepackage{stfloats}
\usepackage{float}
\usepackage{graphicx}
\usepackage{cite}
\usepackage{xcolor}
\usepackage{subfigure}

\makeatletter
\def\blfootnote{\xdef\@thefnmark{}\@footnotetext}
\makeatother

\begin{document}
	
		\title{\huge{Performance Analysis of Fluid Reconfigurable Intelligent Surface over Covert Communications }} 
\author{
	\IEEEauthorblockN{Farshad Rostami Ghadi\textsuperscript{†}, Masoud Kaveh\textsuperscript{‡}, Hanjiang Hong\textsuperscript{*}, Kai-Kit Wong\textsuperscript{*},\\ Riku Jantti\textsuperscript{‡}, and F. Javier Lopez-Martinez\textsuperscript{†}}
	\IEEEauthorblockA{ 
		Department of Signal Theory, Networking and Communications, University of Granada, Granada, Spain.\\}
	\IEEEauthorblockA{\textsuperscript{‡}Department of Information and Communication Engineering, Aalto University, Espoo, Finland. \\}
	\IEEEauthorblockA{\textsuperscript{*}Department of Electronic and Electrical Engineering, University College London, London, United Kingdom.}
	Emails: $\rm \{f.rostami, fjlm\}@ugr.es$; $\rm \{masoud.kaveh, riku.jantti\}@aalto.fi$; $\rm \{hanjiang.hong, kai-kit.wong\}@ucl.ac.uk$
}
	\maketitle
	\begin{abstract}
This paper investigates the impact of the recently proposed concept of fluid reconfigurable intelligent surfaces (FRIS) on covert communications. Specifically, we consider a communication scenario where a legitimate transmitter aims to covertly deliver information to its intended receiver through a planar FRIS, while an adversary attempts to detect whether any transmission is occurring. In this context, we analyze the false alarm (FA) and missed detection (MD) probabilities, and derive a closed-form expression for the covertness outage probability (COP). Furthermore, the success probability is characterized under the optimal detection threshold, providing new insights into the trade-off between covertness and reliable transmission. Numerical results reveal that FRIS provides a clear advantage over fixed-position RIS at low-to-moderate transmit powers by improving reliability and enhancing covertness, while at very high power levels, fixed-position RIS may sustain slightly higher success probability due to reduced leakage toward the adversary.
	\end{abstract}
	%\vspace{-3.5ex}
	\maketitle
	%\blfootnote{\noindent Copyright (c) 2015 IEEE. Personal use of this material is permitted. However, permission to use this material for any other purposes must be obtained from the IEEE by sending a request to pubs-permissions@ieee.org.} 
%	\blfootnote{Manuscript received January 25, 2021; revised XXX. The review of this paper was coordinated by XXXX.} 
%		\blfootnote{This work has been funded in ..}
%	 	\blfootnote{\noindent The authors are with the .. (e-mail: $\rm$.}
	
%	\blfootnote{Digital Object Identifier 10.1109/XXX.2021.XXXXXXX}
	%\IEEEpeerreviewmaketitle
	\vspace{0mm}
	\section{Introduction}\label{sec-intro}
Reconfigurable intelligent surfaces (RISs) have been widely investigated as a low-cost and energy-efficient solution to reshape the wireless propagation environment by coherently controlling electromagnetic reflections through a large number of passive elements~\cite{basar2019}. Despite their promising features, conventional RISs suffer from several key limitations: $i)$ the multiplicative path-loss effect due to the cascaded transmitter-RIS-receiver link, which often results in severe attenuation, and $ii)$ their rigid geometry, which prevents effective exploitation of spatial diversity in environments where physical space is abundant~\cite{wei2021channel}.

To address these limitations, the concept of fluid reconfigurable intelligent surface (FRIS)  \cite{salem2025first} and fluid integrated reflecting and emitting surface (FIRES) \cite{ghadi2025fires} have recently emerged, inspired by the fluid antenna system (FAS)~\cite{wong2020fluid,wong2020performance,new2025tot,ghadi2025ris}. Unlike fixed-position RISs, where elements are fixed in position, FRIS enables position reconfigurability by allowing each reflecting element to dynamically switch its active port within a dense preset grid distributed across the surface. This quasi-continuous spatial reconfiguration provides additional degrees of freedom (DoFs) beyond phase control, enabling adaptive element selection, and stronger constructive alignment toward intended receivers. Early studies have demonstrated that FRIS achieves substantial improvements in outage probability, secrecy rate, and spectral efficiency compared to fixed-position RIS \cite{xiao2025fluid,rostami2025fris,xiao2025beam}. These results highlight FRIS as a strong candidate for future wireless networks, where highly adaptive and robust communications are essential.

Meanwhile, covert communication has gained increasing attention as a new physical layer security (PLS) paradigm, aiming not only to protect information but also to conceal the very existence of transmission from an adversarial warden~\cite{shahzad2017covert,kim2022covert,ghadi2025cov}. Covert techniques are particularly relevant in dense wireless environments, such as vehicular and Internet of Things (IoT) networks, where the risk of detection is high and conventional cryptographic or PLS approaches may be insufficient. Although RIS-assisted covert communications have been studied, the potential of FRIS in this context remains unexplored.

Motivated by this gap, this paper provides the first study of FRIS-aided covert communications. In this regard, we consider a covert communication system where a legitimate transmitter communicates with a legitimate receiver through a planar FRIS, while an adversary attempts to detect the transmission. For this model, we develop a tractable analytical framework to characterize the outage probability (OP), covertness outage probability (COP), and success probability, thereby capturing the fundamental trade-off between reliability and covertness. Closed-form expressions are derived using moment-matching techniques, which enable accurate characterization of the system behavior under FRIS reconfigurability. The analytical and numerical results consistently demonstrate that FRIS can significantly enhance covert communication performance compared to fixed-position RIS at low-to-moderate transmit powers, owing to its adaptive element-selection capability, while at very high power levels the performance gap between FRIS and RIS becomes less pronounced.
\begin{figure}[!t]
	\centering
	\includegraphics[width=0.9\columnwidth]{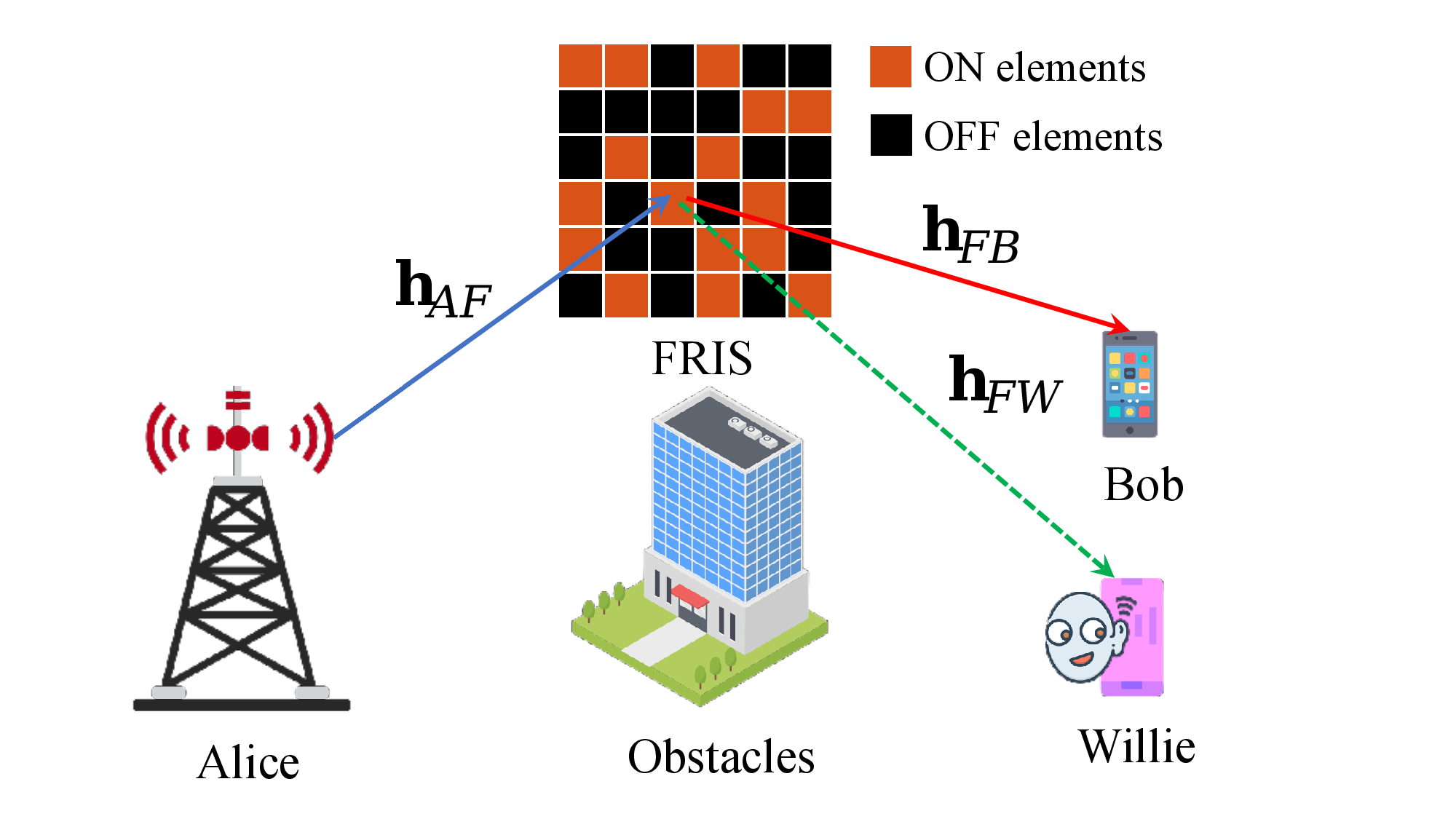}
	\caption{The considered FRIS-FRIS-aided covert communication.}\vspace{0cm}
	\label{fig_model}
\end{figure}
	\section{System Model}\label{sec-sys}
	\subsection{Channel Model}
	We consider a wireless FRIS-assisted covert communication system as presented in Fig. \ref{fig_model}, in which a legitimate transmitter (Alice) intends to covertly deliver information to a legitimate receiver (Bob) via a FRIS, while an adversary (Willie) attempts to detect the presence of the hidden transmission. For notation simplicity, we use $A$, $B$, and $W$ define Alice, Bob and Willie, respectively. All nodes are assumed to be equipped with a single fixed-position antenna, where the legitimate users operate in half-duplex mode. Furthermore, the direct communication links from Alice to both Bob and Willie are assumed to be blocked by obstacles. The considered FRIS consists of $M = M_x \times M_z$ reflective elements, each operating in one of two discrete states of ON or OFF, where $M_x$ and $M_z$ specify the element counts along the horizontal ($x$-axis) and vertical ($z$-axis) directions, respectively. In the ON state, an element interacts with the incident electromagnetic wave, thereby enabling the intended system functionality through wave manipulation. Conversely, in the OFF state, the element is terminated by a matched load, which suppresses interaction with the incoming signal and avoids any alteration of the wavefront. The reflecting elements are uniformly arranged over a planar aperture of size $W = W_x \lambda \times W_z \lambda$, where $\lambda$ denotes the carrier wavelength, and $W_x$ and $W_z$ denote the corresponding aperture dimensions normalized to the wavelength.
	
	FRIS elements are placed in close proximity, which necessitates accounting for spatial correlation among different element positions. To characterize this effect, the Jakes’ model is employed to represent the correlation coefficient between any two elements $i$ and $j$ under rich scattering conditions, expressed as $\mu_{i,j} = \mathcal{J}_0\left(\tfrac{2\pi d_{i,j}}{\lambda}\right)$, where $\mathcal{J}_0(\cdot)$ is the zero-order Bessel function of the first kind and $d_{i,j}$ denotes the inter-element distance, i.e., 
	\begin{multline}
		d_{i,j}=\sqrt{d_x^2\left(\widetilde{M}_i-\widetilde{M}_j\right)^2
		+d_z^2\left(\left\lfloor\frac{i}{M_x}\right\rfloor-\left\lfloor\frac{j}{M_x}\right\rfloor \right)^2},
	\end{multline}
	where $\widetilde{M}_i = \mathrm{mod}\left(i,M_x\right)$, $\widetilde{M}_j=\mathrm{mod}\left(j,M_x\right)$, and $d_x=\frac{W_x\lambda}{M_x}$ and $d_z=\frac{W_z\lambda}{M_z}$ denote the physical inter-element spacing in each FRIS row and column, respectively. Thus, the spatial correlation matrix is defined as $\mathbf{J} = \left[\mu_{i,j}\right]_{M\times M}$.
\subsection{Covert Transmission Scheme and Detection Model}
Alice employs a power control (PC) strategy, where her transmit power $P_A$ is adjusted such that the received signal remains embedded within the background noise, thereby enhancing covertness. Willie acts as a warden and employs a likelihood ratio test (LRT) \cite{sobers2017covert} to determine whether Alice is active. He forms the empirical average received power $\overline{P}_W$ over an observation interval and compares it to a decision threshold $\zeta$. A transmission is declared (hypothesis $\mathcal{H}_1$) if $\overline{P}_W \ge \zeta$, and silence is declared (hypothesis $\mathcal{H}_0$) otherwise. This can be compactly expressed as $
\overline{P}_W \underset{\mathcal{H}_0}{\overset{\mathcal{H}_1}{\gtrless}} \zeta.
$.

Two types of conditional decision errors may occur: 
\emph{(i)} a false alarm (FA), where Willie declares $\mathcal{H}_1$ under $\mathcal{H}_0$, and 
\emph{(ii)} a missed detection (MD), where Willie declares $\mathcal{H}_0$ under $\mathcal{H}_1$. 
The corresponding probabilities are denoted by $P_{\mathrm{FA}}$ and $P_{\mathrm{MD}}$, respectively. 

In covert communication, covertness is achieved when Willie's overall ability to correctly detect Alice’s activity is sufficiently low. Therefore, FA and MD are treated through their \emph{unconditional} contribution to Willie's detection performance, as detailed in the subsequent COP analysis.
\subsection{Signal Model}
When Alice is active, she transmits a vector of $K$ symbols denoted by $\mathbf{x}$, where each entry $\mathbf{x}[k]$, $k=1,\dots,K$, satisfies a unit-power constraint such that $\mathbb{E}{|x[k]|^2} = 1$. In contrast, when Alice remains silent, Willie only observes noise. Accordingly, assuming that $M_O$ elements will be turned ON to modulate the incident signal, the received signal at Willie is expressed as
	\begin{align}\label{eq-yw}
	y_W [k] = 
	\begin{cases} 
		\sqrt{P_AL_{AF}L_{FW}} H_{W} x[k] + n_W[k], &  \mathcal{H}_1 \\
		n_W[k],  & \mathcal{H}_0
	\end{cases},
\end{align}
where $L_{u} = \rho_0 d_u^{-\alpha}$, for $u\in\left\{AF, FW\right\}$ indicates the large-scale path-loss in which $\rho_0$ is the reference gain at one meter, $\alpha$ is the path-loss exponent, and $d_u$ denotes the distance of the corresponding links $u$. Additionally, $n_W[k]\sim\mathcal{CN}\left(0,\sigma^2_W\right)$ are i.i.d. receiver noise samples and $H_W$ denotes the end-to-end channel received at Willie through FRIS, which is given by
\begin{align}
H_W = \mathbf{h}_{FW}^\mathsf{H}\mathbf{J}^{\frac{1}{2}}\mathbf{S}_{M_O}^\mathsf{T}\mathbf{\Phi}\mathbf{S}_{M_O}\mathbf{J}^{\frac{1}{2}}\mathbf{h}_{AF}, 
\end{align}
where $\mathbf{h}_u\in{\mathbb{C}}^{M\times1}$ represents the small-scale fading for the Alice-to-FRIS or FRIS-to-Willie links, i.e., $\mathbf{h}_u\sim\mathcal{CN}\left(0,\mathbf{I}_M\right)$. The diagonal matrix $\mathbf{\Phi}=\mathrm{diag}\left(\left[\mathrm{e}^{j\phi_1}, \dots, \mathrm{e}^{j\phi_M}\right]\right)\in\mathbb{C}^{M\times M}$ includes the adjustable phases of the reflecting elements of the FRIS and $\mathbf{S}^T_{M_O}=\left[\mathbf{s}_1,\dots,\mathbf{s}_{M_O}\right]\in\mathbb{R}^{M\times M_O}$ is the element selection matrix, in which $\mathbf{s}_{m_O}$, for $m_O\in\mathcal{M}_O=\left\{1,\dots,M_O\right\}$, indicates one of the columns of the $M\times M$ identity matrix\footnote{In FRIS, only a subset of meta-elements is activated at a time. This behavior is modeled through the binary selection matrix $\mathbf{S}_{M_O}$, which projects the full aperture of size $M$ onto $M_O$ active elements while ensuring orthogonality. The model captures the fluid antenna principle, where the active elements are dynamically chosen to maximize channel gain, with ideal and fully controllable phase shifts assumed.}. 

Therefore, for detection, Willie evaluates the average received power, given by
	\begin{align}\label{eq-pw}
	\overline{P}_W= \frac{1}{K} \sum_{k=1}^{K} \left|y_W\left[k\right]\right|^2=	\begin{cases}
		P_AL_{AF}L_{FW}g_{W}+\sigma^2_W, &  \mathcal{H}_1\\
		\sigma^2_W, & \mathcal{H}_0
	\end{cases},
\end{align}
where $g_{W} = |h_W|^2$ denotes the channel power gain. Equation \eqref{eq-pw} follows from the strong law of large numbers, applied to \eqref{eq-yw} as $K \to \infty$. This enables Willie to reliably estimate the received signal power and subsequently decide on the presence or absence of Alice's transmission by comparing it with the decision threshold.
\section{Performance Analysis}
In this section, we derive analytical expressions for the COP, OP, and success probability to evaluate the performance of the proposed FRIS-aided covert communication. 

\subsection{Covertness Outage Probability}
In covert communication, a \emph{covertness outage} occurs whenever Willie successfully detects Alice's activity \cite{sobers2017covert}. Since the FA and MD probabilities are conditioned on different hypotheses, the unconditional probability that Willie detects correctly is given by
\begin{align}
	P_{\mathrm{DET}} = p_0 (1-P_{\mathrm{FA}}) + p_1 (1-P_{\mathrm{MD}}),
\end{align}
where $p_0 = P(H_0)$ and $p_1 = P(H_1)$ denote the prior probabilities of the hypotheses. 
Accordingly, we define the COP as
\begin{align}
	\label{eq-cop-def}
	P_{\mathrm{CO}} = P_{\mathrm{DET}}.
\end{align}
\subsubsection{Missed Detection Probability}
A MD occurs when Alice transmits while Willie's test incorrectly decides $H_0$. This happens whenever the observed average received power $\overline{P}_W$ does not exceed the threshold $\zeta$. Thus,
\begin{align}
	P_\mathrm{MD} &= \Pr \left( P_A L_{AF} L_{FW} g_W + \sigma_W^2 \leq \zeta \right) \\
	&= \Pr\left(g_W \le \frac{\zeta - \sigma_W^2}{P_A L_{AF} L_{FW}}\right) \\
	&= F_{g_W}\!\left( \frac{\zeta - \sigma_W^2}{P_A L_{AF} L_{FW}} \right) \label{eq-fgw} \\
	&= 
	\begin{cases}
		F_{g_W}(\eta), & \zeta > \sigma_W^2 \\
		0, & \zeta \le \sigma_W^2
	\end{cases},
\end{align}
where $\eta = \frac{\zeta - \sigma_W^2}{P_A L_{AF} L_{FW}}$. Using the Gamma approximation of $g_W$ \cite[Lemma~1]{rostami2025fris}, the CDF of $g_W$ is given by
\begin{align}
	F_{g_W}(g) = \frac{1}{\Gamma(\kappa)} \Upsilon\!\left(\kappa, \frac{g}{\theta}\right),
\end{align}
with $\kappa = \frac{\left( \mathrm{tr}(\widetilde{\mathbf{J}}^2) \right)^2}{\mathrm{tr}(\widetilde{\mathbf{J}}^4)}$, 
$\theta = \frac{\mathrm{tr}(\widetilde{\mathbf{J}}^4)}{\mathrm{tr}(\widetilde{\mathbf{J}}^2)}$, 
and $\widetilde{\mathbf{J}} = \mathbf{S}_{M_O} \mathbf{J} \mathbf{S}_{M_O}^T$.
Thus,
\begin{align}\label{eq-md}
	P_\mathrm{MD} = 
	\begin{cases}
		\frac{1}{\Gamma(\kappa)} \Upsilon\!\left(\kappa,\frac{\eta}{\theta}\right), & \zeta > \sigma_W^2 \\
		0, & \zeta \le \sigma_W^2
	\end{cases}.
\end{align}

\subsubsection{False Alarm Probability}
A FA occurs when Willie decides $H_1$ under $H_0$. When Alice is silent, 
$\overline{P}_W = \sigma_W^2$, so \cite{digham2007on}
\begin{align}
	\label{eq-fa}
	P_\mathrm{FA} = \Pr(\sigma_W^2 \ge \zeta) =
	\begin{cases}
		0, & \zeta > \sigma_W^2 \\
		1, & \zeta \le \sigma_W^2
	\end{cases}.
\end{align}

\subsubsection{Closed-Form COP Expression}
Substituting \eqref{eq-fa} and the MD expression \eqref{eq-md} into \eqref{eq-cop-def} yields
\begin{align}
	P_{\mathrm{CO}} =
	\begin{cases}
		p_0+p_1\big(1 - P_{\mathrm{MD}}\big), & \zeta > \sigma_W^2 \\
		p_1, & \zeta \le \sigma_W^2
	\end{cases}.
\end{align}
Since $P_{\mathrm{CO}}$ is minimized as $\zeta$ increases, Willie chooses the smallest feasible threshold
\begin{align}
	\label{eq-zeta}
	\zeta^\ast = \sigma_W^2 + \mu,
\end{align}
for an arbitrarily small $\mu > 0$, which represents a worst-case adversary.

\subsection{Outage Probability}
In the considered FRIS-assisted covert communication, the OP quantifies the likelihood that Alice's transmission to Bob fails due to inadequate channel capacity. Formally, the OP is defined as
$P_{\mathrm{OUT}} = \Pr\left(C_B < R_B \right)$, where $R_B$ is the target data rate and $C_B=\log_2\left(1+\gamma_B\right)$ denotes the instantaneous channel capacity of the Alice to Bob link through FRIS in which $\gamma_B$ is the SNR at Bob. 

When Alice transmits with power $P_A$, the FRIS constructs an equivalent channel $H_{B}$ towards Bob. The corresponding received SNR is given by
\begin{equation}
	\gamma_B=\frac{P_AL_{AF}L_{FB}|\mathbf{h}_{FB}^\mathsf{H}\mathbf{J}^{\frac{1}{2}}\mathbf{S}_{M_O}^\mathsf{T}\mathbf{\Phi}\mathbf{S}_{M_O}\mathbf{J}^{\frac{1}{2}}\mathbf{h}_{AF}|^2}{\sigma^2_B},
\end{equation}
where $L_{FB} = \rho_0 d_{FB}^{-\alpha}$ denotes the large-scale path-loss of the FRIS-to-Bob link and $d_{FB}$ is the corresponding distance. The term $\mathbf{h}_{FB}$ accounts the small-scale fading for the FRIS-to-Bob links, i.e., $\mathbf{h}_{FB}\sim\mathcal{CN}\left(0,\mathbf{I}_M\right)$.

Following the FRIS channel characterization,  $g_B=|H_B|^2$ is well approximated by a Gamma distribution with shape parameter $\kappa$ and scale parameter $\theta$, $g_B\sim\mathrm{ Gamma}\left(\kappa,\theta\right)$. Thus, the OP is derived as
\begin{align}
	P_{\mathrm{OUT}} & = \Pr\left(\log_2\left(1+\gamma_B\right) \leq R_B \right)\\
	&= F_{g_B}\!\left(\frac{(2^{R_B}-1)\sigma_B^2}{P_A L_{AF}L_{FB}}\right)\\
	&=\frac{1}{\Gamma\left(\kappa\right)}\Upsilon\left(\kappa,\frac{\overline{R}_B}{\theta}\right), \label{eq-op}
\end{align}
where $F_{g_B}(g)$ denotes the CDF of the Gamma-distributed equivalent channel gain $g_B$ and $\overline{R}_B = \frac{(2^{R_B}-1)\sigma_B^2}{P_A L_{AF}L_{FB}}$.

\subsection{Success Probability}
A transmission is successful only if Bob reliably decodes the message and Willie fails to detect Alice \cite{he2017on}. Using the unconditional error probability \cite{bash2013lim}
\begin{align}
	\xi = p_0 P_{\mathrm{FA}} + p_1 P_{\mathrm{MD}},
\end{align}
the success probability is defined as
\begin{align}
	\label{eq-suc}
	P_{\mathrm{SUC}} = (1 - P_{\mathrm{OUT}})\, \xi,
\end{align}
where $\xi$ is evaluated at the optimal threshold $\zeta^\ast$ in \eqref{eq-zeta}. This expression jointly captures covertness and reliability without requiring independence between events under $H_0$ and $H_1$.

\begin{figure}[!t]
	\centering
	\includegraphics[width=0.9\columnwidth]{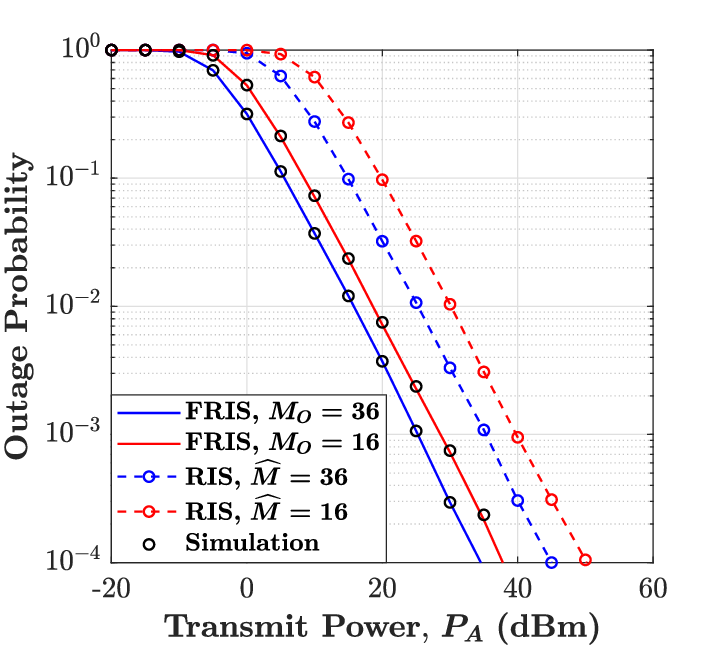}
	\caption{OP versus transmit power $P_A$ for different number of fluid elements $M_O$.}\vspace{0cm}
	\label{fig_op}
\end{figure}
\begin{figure}[!t]
	\centering
	\includegraphics[width=0.9\columnwidth]{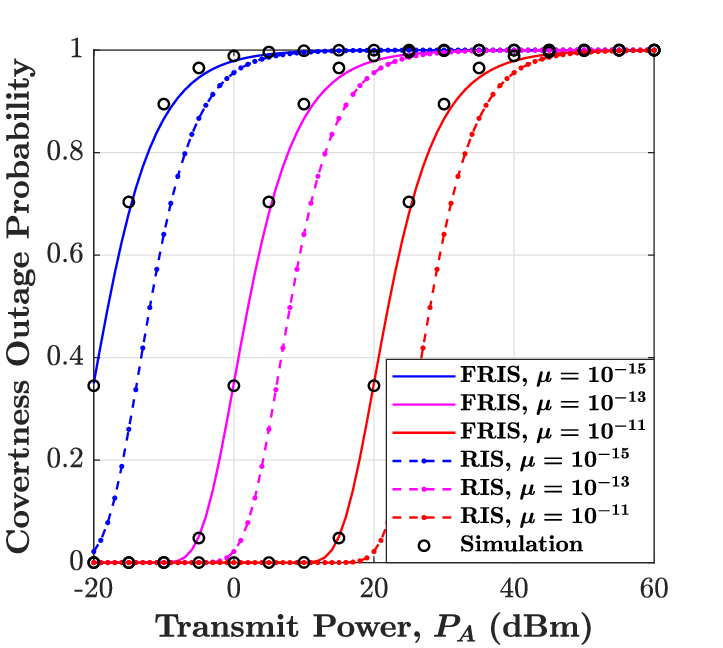}
	\caption{COP versus transmit power $P_A$ for different $\mu$.}\vspace{0cm}
	\label{fig_cop}
\end{figure}
\begin{figure}[!t]
	\centering
	\includegraphics[width=0.9\columnwidth]{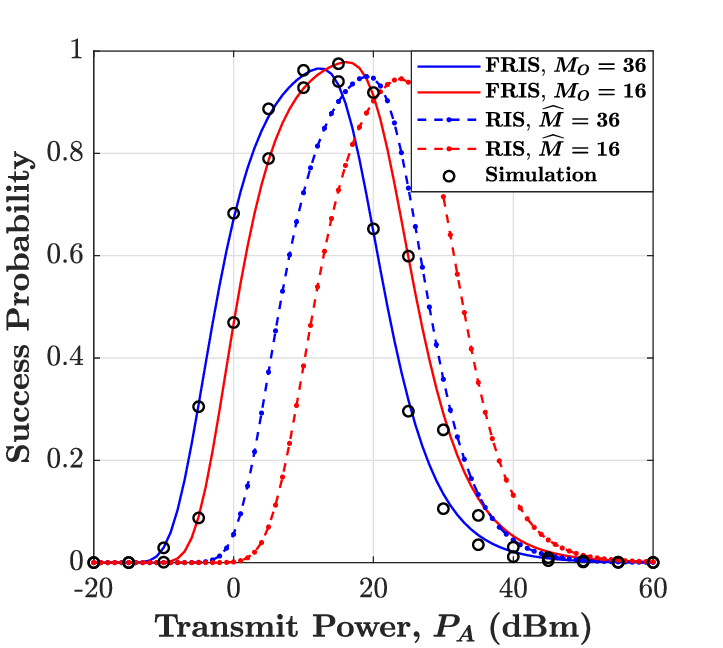}
	\caption{Success probability versus transmit power $P_A$ for different number of fluid elements $M_O$.}\vspace{0cm}
	\label{fig_suc}
\end{figure}
\section{Numerical Results}\label{sec-num}
In this section, we present numerical results to evaluate the performance of the proposed FAS-aided covert communication system. Unless otherwise stated, the system parameters are set as $d_{AF}=50$ m, $d_{FB}=d_{FW}=100$ m, $\alpha=2.1$, $R_B = 0.1$ bits, $\rho_0=1$, $M=(M_x,M_z)=(12,12)$, $M_O=\left\{16,36\right\}$, $W=(W_x,W_z)\lambda^2=(2,2)\lambda^2$, $p_0=p_1=0.5$, and $\sigma^2_B=\sigma^2_W=-90$ dBm. We also assume the carrier frequency $f_c$ at $2.4$ GHz, hence,
$\lambda = 0.125$ m. Furthermore, we benchmark the performance against a fixed-position RIS with optimized phases and the same physical size as the FRIS, where $\widehat{M}$ denotes the number of reflecting elements.

Fig. \ref{fig_op} depicts the OP performance as a function of the transmit power $P_A$ for different numbers of reflecting elements in FRIS and reference RIS, denoted by $M_O$ and $\widehat{M}$, respectively. The close agreement between the simulation and analytical curves confirms the accuracy of the derived expressions. As expected, the OP decreases monotonically with $P_A$, since higher transmit power improves the SNR at Bob. Increasing the number of reflecting elements further enhances reliability, as larger surfaces provide stronger reflected signals and higher diversity gain. It is noteworthy that FRIS achieves a steeper OP reduction compared to RIS for the same number of elements. This performance gain arises from the fluid-based reconfiguration capability of FRIS, which enables adaptive selection of the most favorable subset of elements to maximize the cascaded Alice-FRIS-Bob link. In contrast, reference RIS lacks such flexibility, thereby limiting its effectiveness in mitigating unfavorable fading conditions.

Fig. \ref{fig_cop} shows the covertness outage probability $P_{\rm CO}$ as a function of the transmit power $P_A$ for several threshold offsets $\mu$ (see \ref{eq-zeta}). Consistent with the power-detector model, $P_{\rm CO}$ increases monotonically with $P_A$, since larger transmit power lowers the normalized threshold $\mu/(P_A L_{AF}L_{FB})$, making successful detection by Willie more probable. The parameter $\mu$ shifts the operating point; larger $\mu$ implies a higher threshold and thus delays the onset of outage to higher powers. In the Bob-centric configuration considered here, FRIS curves appear to the left of those for RIS, indicating higher COP at a given $P_A$, which is attributed to element selection/phasing that maximizes Bob’s cascade but can also increase Willie’s effective gain. It is noted that Willie-aware FRIS designs, e.g., penalized selection, would shift the FRIS curves rightward and thereby reduce COP.

Fig. \ref{fig_suc} illustrates the success probability as a function of the transmit power $P_A$ for both FRIS and reference RIS under different numbers of reflecting elements. The success probability exhibits a unimodal trend: at low transmit powers, the link to Bob suffers from severe outage, resulting in negligible success probability. As $P_A$ increases, channel reliability improves while Willie’s detector remains ineffective, driving the success probability to its maximum. However, beyond a certain power level, Willie’s detection becomes more reliable, which significantly reduces the success probability and yields the observed bell-shaped curves. When comparing the two architectures, FRIS consistently achieves higher success probability at lower transmit powers due to its element-selection capability, which strengthens the Alice–Bob channel while suppressing leakage to Willie. At very high transmit powers, however, the FRIS curves decay faster than those of reference RIS, since the latter, optimized toward Bob without active selection, induces less unintentional alignment toward Willie. Furthermore, increasing the number of reflecting elements shifts the curves leftward and raises the peak value, demonstrating that larger surfaces are more power-efficient and provide a higher likelihood of successful covert communication.

\section{Conclusion}\label{sec-con}
This paper studied covert communication in the presence of a FRIS and compared its performance with that of a reference RIS. A system model was developed in which Alice aims to communicate covertly with Bob while an adversary, Willie, attempts to detect the transmission. We derived a closed-form expression for the COP in terms of the FA and MD probabilities. Furthermore, we characterized the success probability under the optimal detection threshold, highlighting the trade-off between covertness and reliable communication. Our numerical results demonstrated that FRIS provides substantial advantages over reference RIS in terms of enhancing reliability and covertness, especially at low-to-moderate transmit power regimes. The element-selection capability of FRIS enables stronger reinforcement of the legitimate Alice-to-Bob channel while suppressing leakage toward Willie, thereby reducing OP and improving the MD performance. However, results also indicated that at very high transmit powers, reference RIS may appear more robust in sustaining success probability, since FRIS’s adaptive focusing can inadvertently enhance the signal leakage to Willie. This insight highlights a key power-dependent trade-off in FRIS-aided covert communication.

\section*{Acknowledgment}
{The work of F. Rostami Ghadi is supported by the European Union's Horizon 2022 Research and Innovation Programme under Marie Skłodowska-Curie Grant No. 101107993. The work of K. K. Wong is supported by the Engineering and Physical Sciences Research Council (EPSRC) under grant EP/W026813/1. The work of M. Kaveh and R. Jäntti has received funding from the SNS JU under the EU’s Horizon Europe Research and Innovation Programme under Grant Agreement No. 101192113 (Ambient-6G). The work of F.J. López-Martínez is supported by grant PID2023-149975OB-I00 (COSTUME) funded by MICIU/AEI/10.13039/501100011033 and FEDER/UE.} 
%\bibliographystyle{IEEEtran}
%\bibliography{refs.bib}

\begin{thebibliography}{1}
	\bibitem{basar2019}
	 E. Basar, M. Di Renzo, J. De Rosny, M. Debbah, M. -S. Alouini and R. Zhang, ``Wireless Communications Through Reconfigurable Intelligent Surfaces," \textit{IEEE Access}, vol. 7, pp. 116753-116773, 2019.
	 
	\bibitem{wei2021channel}
	M. Jian {\em et al.}, ``Reconfigurable intelligent surfaces for wireless communications: Overview of hardware designs, channel models, and estimation techniques,'' {\em Intelligent \& Conv. Netw.}, vol. 3, no. 1, pp. 1--32, 2022.
	
	\bibitem{salem2025first}
	A. Salem, K. K. Wong, G. Alexandropoulos, C.-B. Chae, and R. Murch, ``A first look at the performance enhancement potential of fluid reconfigurable intelligent surface,'' {\em arXiv preprint}, {arXiv:2502.17116v1}, 2025.
	
	\bibitem{ghadi2025fires}
	F. Rostami Ghadi, K. -K. Wong, M. Kaveh, F. J. López-Martínez, C. -B. Chae and G. C. Alexandropoulos, ``FIRES: Fluid Integrated Reflecting and Emitting Surfaces," \textit{IEEE Wirel. Commun. Lett.}, vol. 14, no. 11, pp. 3744-3748, Nov. 2025.

	\bibitem{wong2020fluid}  
	K. K. Wong, A. Shojaeifard, K. F. Tong, and Y. Zhang, ``Fluid antenna systems,” \textit{IEEE Trans. Wirel. Commun.}, vol. 20, no. 3, pp. 1950–1962, Mar. 2020.
	
	\bibitem{wong2020performance}
	K. K. Wong, A. Shojaeifard, K. -F. Tong and Y. Zhang, ``Performance Limits of Fluid Antenna Systems," \emph{IEEE Commun. Lett.}, vol. 24, no. 11, pp. 2469-2472, Nov. 2020.
	
	\bibitem{new2025tot}
	W. K. New {\em et al.}, ``A Tutorial on Fluid Antenna System for 6G Networks: Encompassing Communication Theory, Optimization Methods and Hardware Designs," \emph{IEEE Commun. Surv. Tutor.}, vol. 27, no. 4, pp. 2325-2377, Aug. 2025.
	
	\bibitem{ghadi2025ris}
	F. R. Ghadi, K. K. Wong, W. K. New, H. Xu, R. Murch, and Y. Zhang, ``On performance of RIS-aided fluid antenna systems,” \emph{IEEE Wireless Communications Letters}, Vol. 13, No. 8, pp. 2175-2179, August 2024.
	
	\bibitem{xiao2025fluid}
	H. Xiao {\em et al.}, ``Fluid reconfigurable intelligent surfaces: Joint on-off selection and beamforming with discrete phase shifts,'' {\em IEEE Wirel. Commun. Lett.}, vol. 14, no. 10, pp. 3124-3128, Oct. 2025.
	
	\bibitem{rostami2025fris}
	F. Rostami Ghadi, K. -K. Wong, F. J. López-Martínez, G. C. Alexandropoulos and C. -B. Chae, ``Performance Analysis of Wireless Communication Systems Assisted by Fluid Reconfigurable Intelligent Surfaces," \textit{IEEE Wirel. Commun. Lett.}, {doi: 10.1109/LWC.2025.3608040}, 2025. 
	
	\bibitem{xiao2025beam}
	H. Xiao, X. Hu, K-K. Wong, X. Zhu, H. Hong, and Ch-B. Chae, ``Fluid Reconfigurable Intelligent Surface with Element-Level Pattern Reconfigurability: Beamforming and Pattern Co-Design," {\em arXiv preprint}, {arXiv:2508.09695v1}, 2025.
	
%	\bibitem{new2024tutorial}  
%	W. K. New \textit{et al.}, “A tutorial on fluid antenna system for 6G networks: Encompassing communication theory, optimization methods and hardware designs,” \emph{IEEE Commun. Surv. Tutor.}, 2024.
%	\bibitem{tang2023fas}  
%	B. Tang, H. Xu, K. K. Wong, K. F. Tong, Y. Zhang, and C. B. Chae, “Fluid antenna enabling secret communications,” \emph{IEEE Commun. Lett.}, vol. 27, no. 6, pp. 1491–1495, Jun. 2023.
%	\bibitem{vega2024sop}  
%	J. D. Vega-Sánchez, L. Urquiza-Aguiar, H. R. C. Mora, N. V. O. Garzón, and D. P. M. Osorio, “Fluid antenna system: Secrecy outage probability analysis,” \emph{IEEE Trans. Veh. Technol.}, vol. 73, no. 8, pp. 11458–11469, Aug. 2024.
%	\bibitem{ghadi2024secure}  
%	F. R. Ghadi \textit{et al.}, “Physical layer security over fluid antenna systems: Secrecy performance analysis,” \emph{IEEE Trans. Wireless Commun.}, vol. 23, no. 12, pp. 18201–18213, Dec. 2024. 
%\bibitem{Yao2024covert}  
%J. Yao \textit{et al.}, “FAS for secure and covert communications,” \emph{arXiv preprint}, \url{arXiv:2411.09235},  Nov. 2024.
\bibitem{shahzad2017covert}
K. Shahzad, X. Zhou and S. Yan, ``Covert Communication in Fading Channels under Channel Uncertainty," \emph{2017 IEEE 85th Vehicular Technology Conference (VTC Spring)}, Sydney, NSW, Australia, 2017, pp. 1-5.
\bibitem{kim2022covert}
S. W. Kim and H. Q. Ta, ``Covert Communications Over Multiple Overt Channels," \emph{IEEE Trans. Commun.}, vol. 70, no. 2, pp. 1112-1124, Feb. 2022.
\bibitem{ghadi2025cov}
F. Rostami Ghadi, M. Kaveh, R. Janti, and F. J. Lopez-Martinez ``On Performance of FAS-aided Covert Communications," \emph{2025 IEEE 101st Vehicular Technology Conference (VTC2025-Spring)}, Oslo, Norway, 2025, pp. 1-5.
\bibitem{sobers2017covert}
T. V. Sobers, B. A. Bash, S. Guha, D. Towsley and D. Goeckel, ``Covert Communication in the Presence of an Uninformed Jammer," \textit{IEEE Trans. Wirel. Commun.}, vol. 16, no. 9, pp. 6193-6206, Sept. 2017.

\bibitem{digham2007on}
F. F. Digham, M. -S. Alouini and M. K. Simon, ``On the Energy Detection of Unknown Signals Over Fading Channels," \textit{IEEE Trans. Commun.}, vol. 55, no. 1, pp. 21-24, Jan. 2007.

\bibitem{he2017on}
B. He, S. Yan, X. Zhou and V. K. N. Lau, ``On Covert Communication With Noise Uncertainty," \textit{IEEE Commun. Lett.}, vol. 21, no. 4, pp. 941-944, April 2017.

\bibitem{bash2013lim}
B. A. Bash, D. Goeckel and D. Towsley, ``Limits of Reliable Communication with Low Probability of Detection on AWGN Channels," \textit{IEEE J. Sel. Area. Commun.}, vol. 31, no. 9, pp. 1921-1930, September 2013.

	\end{thebibliography}

\end{document}